\documentstyle[aps,twocolumn,floats,prl]{revtex}

\newcommand{\beq}{\begin{equation}}
\newcommand{\eeq}{\end{equation}}
\def\beqa{\begin{eqnarray}}
\def\eeqa{\end{eqnarray}}
\def\p{\partial}

\def\lap{\lower.5ex\hbox{$\; \buildrel < \over \sim \;$}}
\def\gap{\lower.5ex\hbox{$\; \buildrel > \over \sim \;$}}
\def\lb{\langle}
\def\rb{\rangle}
\def\fv{{\bf f}}
\def\vv{{\bf v}}
\def\rv{{\bf r}}
\def\xv{{\bf x}}

\input psfig

\begin{document}
\title{Dynamics of a Dark Matter Field with a Quartic
Self-Interaction Potential} 
\author{P. J. E. Peebles\\
Joseph Henry Laboratories, Princeton University, Princeton NJ
08544}

\date{\today}
\maketitle
\begin{abstract}
It may prove useful in cosmology to understand the
behavior of the energy distribution in a scalar field that
interacts only with gravity and with itself by a pure quartic
potential, because if such a field existed it would be
gravitationally produced, as a squeezed state, during inflation.
It is known that the mean energy density in such a field after
inflation varies with the expansion of the universe in the
same way as radiation. I show that if the field initially is
close to homogeneous, with energy density contrast 
$|\delta\rho /\rho|\ll 1$ and coherence length $L$, the energy
density fluctuations behave like acoustic oscillations in an
ideal relativistic fluid for a time on the order of
$L/|\delta\rho /\rho |$. This ends with the appearance of features 
that resemble shock waves but interact in a
close to elastic way that reversibly disturbs the energy
distribution. 
\end{abstract}

\section{Introduction}

The dark matter that in standard physics dominates the mass of
the universe might have originated in the same way as ordinary
matter and radiation, in reactions at high redshift, or some or
all of it might have been gravitationally produced by inflation
\cite{PVa}, \cite{PVb}. A simple example of the latter is a
single scalar field $y(\xv ,t)$ with the pure quartic self-interaction
potential 
\beq
V = \lambda y^4/4,
\label{eq:V}
\eeq
minimal coupling to gravity, and no other interactions. 
This could not be the dark matter that is gravitationally bound
to galaxies and clusters of galaxies, but it might have
an observationally interesting effect on structure
formation. A model based on the idea that the primeval
fluctuations in this $y$-field energy distribution broke
homogeneity to seed structure formation is compared to the
measured power spectra of the radiation and matter distributions 
in \cite{HP}. Here I consider some aspects of the
evolution of the $y$-field energy distribution. 

Ford \cite{Ford} and Turner \cite{Turner} show that after 
inflation ends the energy density in a homogeneous field with
the quartic potential in Eq~(\ref{eq:V}) 
scales with the cosmological expansion parameter as 
$\rho _y\propto a(t)^{-4}$,
the same as for radiation. This is important because it means the
$y$-field energy could be large enough when inflation ends to be
a significant perturbation to the energy distribution, yet small
enough thereafter that it does not interfere with the standard
models for light element production at redshift $z\sim 10^{10}$
and structure formation at lower redshift. It may be significant
also that the parameter $\lambda$ in the quartic potential
can be chosen so that eternal inflation gravitationally produces a
potentially interesting value for the energy density $\rho _y$. 
There is the danger noted by Felder, Kofman and Linde \cite{fkl}
that gravitational production can make $\rho _y$ the dominant
energy density, so this field assumes the role of the
inflaton. Peebles and Vilenkin \cite{PVb} show that in eternal
inflation, with the quartic inflaton potential 
$V_\phi =\lambda _\phi \phi ^4/4$, this problem does not arise if
the $y$-field parameter satisfies $\lambda _\phi\ll\lambda$.

Since the mean value of $\rho _y$ after inflation
varies in the like radiation it might not seem
surprising that energy density fluctuations can behave in a
similar way. An early step in this direction was 
taken by Khlopov, Malomed, and Zel'dovich \cite{KMZ}, who note
that the quartic potential with positive $\lambda$ produces a
positive effective pressure. Other steps toward the fluid picture
are reviewed in
Section~II. A linear acoustic wave model for the evolution of
small spatial fluctuations in the field energy density can only
be an approximation: the numerical solution of Khlebnikov \&\
Tkachev \cite{KT} shows the transfer of fluctuations in the
spatial distribution of the energy to higher wavenumbers that
Gruzinov \cite{Gruzinov} remarks can be interpreted as the
relaxation to equipartition of the kinetic, gradient and
potential energies. The purpose of this paper is to find a
description of the onset of the departure from the acoustic wave
model. 

This analysis assumes $\p y/\p t =0$ at the end of
inflation, as follows from gravitational production of $y$, and
the field value is close to homogeneous. Peebles and Vilenkin
\cite{PVb} show this last assumption is a good approximation
across the present Hubble length if $\lambda\lap 0.01$. 
Khlopov, Malomed, and Zel'dovich \cite{KMZ} study the
gravitational effect of the evolution of the energy distribution
in a scalar field with quartic plus quadratic self-interaction
potential, in a WKB approximation. Here I consider a pure quartic
potential, where the effect of the peculiar gravitational
acceleration is likely to be important only near black holes. 
I accordingly use an expanding spacetime that is homogeneous and
isotropic.  The $y$-field is supposed to have been squeezed by
inflation to very large occupation numbers, so it can be treated 
as a classical field. 

\section{Preliminary Remarks}

When the field is exactly homogeneous, the field equation in
Minkowski spacetime can be scaled to a one-dimensional
oscillator,  
\beq
\ddot\eta = -\eta ^3, \qquad -1\leq\eta\leq 1,
\label{eq:eta_eq}
\eeq
where $\dot y =\p y/\p t$, $t$ is proper
world time, and $\eta (t)$ has unit
amplitude. Following Ford \cite{Ford}, one arrives at a virial
relation by  multiplying Eq.~(\ref{eq:eta_eq}) by $\eta$
and averaging over time. The left hand side is 
$\eta\ddot\eta = d(\eta\dot\eta )/dt - \dot\eta ^2$; in 
the average over time large compared to the oscillator
period the second term is dominant, giving \cite{Ford}, 
\cite{Greene}
\beq
\lb\dot\eta ^2\rb = \lb\eta ^4\rb = 1/3,
\label{eq:averages}
\eeq
where the energy is $\dot\eta ^2/2 + \eta ^4/4=1/4$ for unit
amplitude. Greene {\it et al.} \cite{Greene} show
$\eta (t)$ has frequency
\beq
\omega =2\pi ^{3/2}\Gamma (1/4)^{-2} = 0.847\ldots .
\label{eq:omega}
\eeq 
The scaling to potential $\lambda y^4/4$ and amplitude $y_o$ is 
\beq
y = y_o\eta (\lambda ^{1/2}y_ot), \qquad 
\omega = 0.847 \lambda ^{1/2}y_o.
\label{eq:y(t)}
\eeq
Turner \cite{Turner} notes that when the field is spatially
homogeneous, $y=y(t)$, with amplitude $y_o$, the time average of
the stress-energy tensor $T_i^j$ is diagonal with 
$\rho =\dot y^2/2 + \lambda y^4/4 = \lambda \lb y_o^4\rb /4$ and 
$p =\lb\dot y^2/2 - \lambda y^4/4\rb = \lambda \lb y_o^4/12\rb$,
the time averages following from Eq.~(\ref{eq:averages}).
The ratio $p/\rho$ is the same as for an
ideal relativistic fluid. One sees from the
vanishing of the covariant divergence of this diagonal
stress-energy tensor in comoving coordinates that the energy
density scales with the cosmological expansion parameter as
\beq
\rho\propto a(t)^{-4}.
\label{eq:rho_evolution}
\eeq
Other derivations are given by Ford \cite{Ford} and Turner
\cite{Turner}.  

If the field is inhomogeneous the action with quartic potential 
is simplified by changing from proper time $t$ to conformal
time $\tilde t=\int dt/a$ and letting $\tilde y=ay$.
When $\omega\gg\dot a/a$ this brings the action for $\tilde y$ to
the Minkowski form. In this limit a solution 
$\tilde y(\tilde t,{\bf x})$ to the Minkowski problem,
that has fixed mean energy density, produces a solution 
$y(t,{\bf x})$ in which the kinetic, gradient and potential
energy densities all scale as $a(t)^{-4}$ relative to the
Minkowski solution. The mean energy density thus satisfies 
Eq.~(\ref{eq:rho_evolution}) (eg. \cite{PVb}, \cite{SY}). 

One gets some feeling for the evolution of the energy
distribution in an inhomogeneous solution by starting from the
homogeneous solution obtained by a Lorentz transformation of
Eq.~(\ref{eq:y(t)}), 
\beq 
y(\rv ,t) = y_o\eta (\lambda ^{1/2}y_o\gamma (t-\vv\cdot\rv )).
\label{eq:y(t,x)}
\eeq
The time average of the stress-energy tensor $T_{ij}$ in this
solution is the same 
as for an ideal relativistic fluid with streaming velocity $\vv$,
as one sees from the Lorentz transformation from $T_{ij}$ for
$y=y(t)$ or by computing $T_{ij}$ for $y(\rv ,t)$ in
Eq.~(\ref{eq:y(t,x)}). If the field initially is locally well
approximated by 
Eq.~(\ref{eq:y(t,x)}), with amplitude $y_o$ and velocity $\vv$
that vary slowly with position, then $T_{ij}$ is close to 
the stress-energy tensor of an inhomogeneous relativistic ideal
fluid, and the vanishing divergence of 
$T_{ij}$ tells us the energy starts to redistribute itself 
in the manner of a relativistic fluid. (This was pointed out to
me by Vilenkin \cite{Vilenkin}). The analysis in the next section
is meant to find indications of how long this fluid approximation
might last. 

The discussion conveniently begins with yet
another demonstration of the evolution of the mean energy density
in Eq.~(\ref{eq:rho_evolution}) when the field is inhomogeneous.  
The wave equation is
\beq
{\p ^2y\over\p t^2} + 3{\dot a\over a}{\p y\over\p t} = 
{\nabla ^2 y\over a^2} - \lambda y^3,
\label{eq:wave_eq}
\eeq
and the energy density and flux are
\beq
\rho = \dot y^2/2 + (\nabla y)^2/2a^2 + \lambda y^4/4, \qquad
{\bf f} = - \dot y\nabla y /a,
\label{eq:rho}
\eeq
where the space derivatives are with respect to coordinate distance
$x$ and $a dx$ is a proper length. The energy conservation equation is  
\beq
\p\rho /\p t +\nabla\cdot {\bf }f/a = 
	-[3\dot y^2 + (\nabla y)^2/a^2]\dot a/a.
\label{eq:rho_cons}
\eeq
As in Eq.~(\ref{eq:averages}), the result of multiplying
Eq. (\ref{eq:wave_eq}) by $y$ and averaging over space and over a
time interval large compared to the period $\omega ^{-1}$ of
oscillation of $y$ and small compared to the expansion time
$a/\dot a$ is the virial relation 
\beq
\lb\dot y^2\rb = \lb (\nabla y)^2\rb /a^2 + \lambda \lb y^4\rb .
\label{eq:virial}
\eeq
This brings the mean energy density to
\beq
\lb\rho\rb = 3\lb\dot y^2\rb /4+ \lb\nabla y^2\rb/4a^2 .
\label{eq:mean_rho}
\eeq
The space average of Eq.~(\ref{eq:rho_cons}) eliminates the
energy flux density, and with Eq.~(\ref{eq:mean_rho}) we get 
$\lb\rho\rb\propto a(t)^{-4}$.

%For yet another way to look at this result consider the space
%part of the stress-energy tensor of $y$. The diagonal part in the
%direction of $x^1$ is
%\beq
%T_{11}=(\p y/\p x^1)^2/a^2 + \dot y^2 -(\nabla y^2)^2/a^2
% -\lambda y^4/4.
%\label{eq:T11}
%\eeq
%If the field fluctuations are isotropic the mean of the first
%term on the right hand side is $(\nabla y^2)^2$/3, 
%and the space average of Eq.~(\ref{eq:T11}) is the effective
%pressure  
%\beq
%p=\lb\dot y^2\rb /4+ \lb\nabla y^2\rb/12a^2 .
%\eeq
%This is one third of the energy density, consistent with
%Eq.~(\ref{eq:rho_evolution}). 

If the sum of the kinetic and gradient energy densities
is large compared to the potential energy then $y$ has the
familiar properties of a free massless field: initial 
large-scale fluctuations in the energy distribution grow smoother
as in free streaming of a gas of relativistic particles.  
In the situation to be discussed next the initially small
gradient energy density grows larger in patches and may
eventually become comparable to the kinetic and 
potential energies. In this case large-scale density fluctuations
would evolve like a relativistic gas 
of massless particles with mean free path comparable to a
characteristic de~Broglie wavelength. 

The model presented in the next section is meant to describe the
onset of this cascade to larger wavenumbers, when
the field  gradient energy density is small compared to the
kinetic plus potential energy densities. The examples
in Section IV show that
numerical solutions to this model are in satisfactory agreement
with the energy distributions from numerical solutions of the  
wave equation. Both demonstrate the failure of the linear
acoustic wave model for the behavior of fluctuations in the
energy distribution, in the development of features that 
resemble shock waves except that they pass each other in a
nearly free way. 

\section{The Weak gradient model}

Since the period of oscillation of $y$ scales inversely with 
the amplitude (Eq.~\ref{eq:y(t)}), an initial gradient of the
amplitude produces a growing gradient of the phase of the
oscillation. This makes the 
field an oscillating function of position, as in
Eq.~(\ref{eq:y(t,x)}), and produces energy flux  
{\bf f} that tends to carry energy from maxima toward minima of
the distribution. When the field gradient is on the order of 
$\nabla y\sim v\dot y$ the streaming velocity in this
rearrangement of the energy is $\sim f/\rho\sim v$.
In acoustic waves in a relativistic fluid $v$ is small when the
density fluctuations are small, $|\delta\rho /\rho |\ll 1$. 
Following this condition, the weak gradient model (WGM) for the
energy distribution in the $y$-field assumes two inequalities.
First, the period $\sim\omega ^{-1}$ of oscillation of the field
is much shorter than the wavelength $\sim (v\omega )^{-1}$ of
spatial oscillations, because $v\ll 1$.
Second, the wavelength of field oscillations is much shorter 
than the coherence length $L$ of the energy distribution. 
That is, the WGM assumes $\omega\gg\omega v\gg 1/L$.

The goal is to study the redistribution of energy on the
scale of $L$, not the more rapid oscillations 
of $y$ in space and time, so the energy density and flux are
smoothed over a window in spacetime that is large compared to the
latter and small compared to $L$. The virial relation in
Eq~(\ref{eq:virial}) applies to the averages over this
intermediate scale, and the smoothed energy
distribution accordingly satisfies Eq.~(\ref{eq:mean_rho}).
The same smoothing of the energy Eq.~(\ref{eq:rho_cons}) with
Eq.~(\ref{eq:mean_rho}) yields
\beq
\p\lb\rho\rb /\p t +\nabla\cdot\lb\fv\rb /a =- 4\lb\rho\rb\dot a/a .
\label{eq:wgaet}
\eeq
The angular brackets mean the result of this smoothing.

Eq.~(\ref{eq:wgaet}) is the first relation in the WGM.  
The second follows by writing out the time derivative of  
the energy flux density (Eq.~\ref{eq:rho}) and using the wave
Eq.~(\ref{eq:wave_eq}):
\beq
{\p\fv\over\p t} + 4\fv {\dot a\over a} = 
{1\over a}\nabla \left({1\over 4}\lambda y^4 
- {1\over 2}\dot y^2)\right) - {1\over a^3}\nabla y\nabla ^2y. 
\label{eq:1}
\eeq
Eq~(\ref{eq:mean_rho}) brings the mean of the
first term on the right hand side to 
\beq
\lb\lambda y^4/4 -\dot y^2/2\rb = -\lb\rho\rb /3 
	-\lb (\nabla y)^2\rb /(6a^2).
\label{eq:3}
\eeq
The last term is simplified by noting that the energy flux 
density satisfies 
\beq
\fv \nabla\cdot\fv = {1\over a^2}\dot y^2\nabla y\nabla ^2y
- {1\over 2a}\fv {\p (\nabla y)^2\over\p t} .
\label{eq:2}
\eeq
Eq.~(\ref{eq:2}) and the second term on the right hand side of
Eq.~(\ref{eq:3}) will serve as small corrections to a wave  
equation for $\rho$ and $\fv$, so I approximate these parts by 
their forms for a homogeneous energy distribution with streaming
velocity $\vv$. Eq.~(\ref{eq:y(t,x)}) with $v\ll 1$
says $\nabla y/a \cong -\vv\dot y$. 
With $\lb\dot y^2\rb\cong 4\lb\rho\rb /3$ the energy flux density 
is $\lb\fv\rb \cong 4\lb\rho\rb\vv /3$. 
Thus in the last term in Eq.~(\ref{eq:3}) we have
\beq
\lb (\nabla y)^2\rb/a^2\cong (3/4)f^2/\rho .
\eeq
The last term on the right hand side
of Eq.~(\ref{eq:2}) is on the order 
of $v^2$ times the left hand side, and so is discarded. 
In these approximations Eq.~(\ref{eq:1}) becomes
\beq
{\p\fv \over\p t} + 4{\dot a\over a}\fv 
= -{1\over 3a}\nabla\rho - {1\over 8a\rho }\nabla f^2
- {3\over 4\rho a}\fv \nabla\cdot\fv .
\label{eq:5}
\eeq
Here and below I drop the brackets that indicate the smoothing
of the energy density and flux. 

The conformal transformation that
eliminated the expansion of the universe in the action has 
the same effect here. The transformations are 
$\tilde\rho = a^4\rho$, $\tilde\fv = a^4\fv$, $d\tilde t=dt/a$.
Without the tildes, Eqs.~(\ref{eq:wgaet}) and (\ref{eq:5}) in the new
variables are 
\beqa
{\p\rho\over\p t} &=& -\nabla\cdot\fv ,  \nonumber \\ 
{\p\fv\over\p t} &=& -{1\over 3}\nabla\rho - 
{1\over 8\rho }\nabla f^2 - {3\over 4\rho}\fv\nabla\cdot\fv .
\label{eq:wga3}
\eeqa
For motion in one space dimension this becomes 
\beq
{\p\rho\over\p t} = -{\p f\over\p x},\qquad 
{\p f\over\p t} = -{1\over 3}{\p\rho\over\p x} 
-{f\over\rho }{\p f\over\p x},
\label{eq:wga}
\eeq
with the wave equation
\beq
{\p ^2\rho\over\p t^2} = {1\over 3}{\p ^2\rho\over\p x^2}
+{1\over 2\rho }{\p^2f^2\over\p x^2}.
\eeq

Eq.~(\ref{eq:wga3}) is the weak gradient model (WGM) for the
evolution of the field energy density smoothed over scales
small compared to the coherence length of the energy distribution
and large compared to the wavelength and period of oscillation of
the field. This is a computation in second order in the 
perturbation from a static homogeneous energy distribution. 

The first part of the WGM in Eq.~(\ref{eq:wga3}) just expresses
energy conservation after the transformation to the scaled field
in Minkowski spacetime. The second part can be compared to the
behavior of an 
ideal fluid with rest energy density $\rho _r$ and pressure
$p_r=\rho _r/3$. The components of the stress-energy tensor 
$T^{ij}$ for this fluid are
\beqa
T^{00} &=& \rho = (4\gamma ^2 - 1)\rho _r/3, \nonumber \\ 
T^{0\alpha}&=& f^\alpha = 4\rho_r\gamma^2v^\alpha /3, \\
T^{\alpha\beta} &=& 
(\delta ^{\alpha\beta } + 4\gamma ^2v^\alpha v^\beta )\rho _r/3.\nonumber
\eeqa
Here $\alpha$ labels the spatial Cartesian coordinates.
The time part of the divergence of $T^{ij}$ is the first part of
Eq.~(\ref{eq:wga3}). The space part of the divergence, in second
order perturbation theory, is
\beq
{\p\fv\over\p t} = -{1\over 3}\nabla\rho + 
{1\over 4\rho }\nabla f^2 - {3\over 4\rho}\fv (\nabla\cdot\fv )
-{3\over 4\rho }(\fv\cdot\nabla )\fv .
\label{eq:ideal_fluid}
\eeq
If the flow is irrotational in first order perturbation theory,
we can write $(\fv\cdot\nabla )\fv =\nabla f^2/2$ in the last
term. In this case Eq.~(\ref{eq:ideal_fluid}) is the same as the
second part of Eq.~(\ref{eq:wga3}). 

Since the evolution of the field energy distribution is the same
as that of an irrotational relativistic fluid through second
order in the perturbation from the static solution, the growing
departure from the linear acoustic model may be expected to
resemble the development of shock waves. One sees 
this in the WGM by considering solutions to Eq.~({\ref{eq:wga})
in second order perturbation theory. A standing wave solution is 
\beqa
\rho &=& \rho _b[ 1 + A\cos kx\, \cos\Omega t \nonumber \\ 
&+& A^2\cos 2kx\, (1-\cos 2\Omega t - \Omega t\sin 2\Omega t )/8
+ \ldots  ] ,
\label{eq:wga_standing}
\eeqa
where $\rho _b$ is the mean energy density and 
$\Omega =k/\sqrt{3}$. A travelling wave solution is
\beqa
\rho &=& \rho _b [ 1 + A\sin k(x - t/\sqrt{3}) \nonumber \\ 
&-& (\Omega tA^2/4)\sin 2k(x - t/\sqrt{3})
+ \ldots ].
\label{eq:wga_travelling}
\eeqa
The nonlinear correction term in Eq.~(\ref{eq:wga_travelling}) makes
the leading part of a maximum of the energy distribution  
steeper than the trailing part, as in the development of a shock
wave in an ideal fluid. In the standing wave solution the first
nonlinear correction term vanishes when the leading term
vanishes or is at a stationary point; at other times the
correction term makes the  
extremum of the energy distribution more prominent and narrower.

\section{Numerical Solutions}

More detailed illustrations of the evolution of the energy
distribution, and tests of the
WGM, are presented in Figures~1 to~5. The solid curves are
computed from numerical integration of the
wave  Eq.~(\ref{eq:wave_eq}), and the dashed curves are from
numerical solutions to the WGM in Eq.~(\ref{eq:wga}).
In these examples the initial field amplitude is almost 
constant at $y_o$. When the length and time unit is 
$(\lambda ^{1/2}y_o)^{-1}$, and the mean of $y$ is scaled to unit 
amplitude, the field equation is  
\beq
\p ^2 y /\p t^2 = \nabla ^2 y -  y^3,
\label{eq:scaled_wave_eq}
\eeq
and the initial energy density is close to 
$\rho = 1/4$. 

The numerical integration of the WGM uses the
density as a function of position, rather than its mean value, in
the denominator of Eq.~(\ref{eq:wga3}), but that does not much
matter because the energy density fluctuations remain small.
The numerical solutions are periodic in the space intervals in
the figures. 
The second spatial derivative is approximated as
$y_i^{\prime\prime } =(y_{i+1}-2y_i+y_{i-1})/\delta _x^2$,
where $\delta _x$ is the spatial coordinate interval, with the
usual extension to two spatial dimensions. The time integration
uses $y(t+\delta _t) = y(t) + 0.5\delta _t[\dot y(t+\delta t) +
\dot y(t)]$, where $\dot y(t+\delta t)$ is computed by iteration.

The initial condition in Fig.~1 for a standing wave in one
dimension is $\p y/\p t=0$ with 
\beq
y = 1 + B\sin (kx).
\label{eq:53}
\eeq
The amplitude of the initial field fluctuation is $B =0.0125$, so
the maximum initial density contrast is 
$\delta\rho /\rho = A\simeq 4B = 0.05$. The wavenumber is
$k=0.01\lambda ^{1/2}y_o$, or $k=0.01$ in the units of
Eq.~(\ref{eq:scaled_wave_eq}). The labels in the figure 
are the times from the initial condition in Eq.~(\ref{eq:53}) in 
units of the period $T=2\pi\sqrt{3}/k$ of oscillation of the
energy distribution in a linear acoustic wave model. 
The numerical solutions use 700 time steps per period of
oscillation of the field value at the mean amplitude $y_o$, or
about $1\times 10^5$ time steps in the acoustic oscillation
period $T$. Increasing the size of the 
numerical time step by a factor of two or four has little effect
on the energy distribution at the time 
$20T$ shown in the figure. At $40T$ the numerical results
are sensitive to the time step; exploration of this stage of the   
evolution would require a more careful
computation than has been attempted here. There are
350 space steps in the periodic boundary condition. If the
coordinate interval is smaller than this, numerical
instability spoils the integration before $20T$. 

The wavelengths of the small-scale fluctuations behind the shock-like
features in Fig.~1 vary with the coordinate interval $\delta _x$;
the value of the small-scale wavelength is a numerical artifact. I expect 
that in the limit of a continuous field the small-scale oscillations
would approach the wavelength set by the period of oscillation of the
field, since that and $\delta _x$ 
are the only relevant length scales in the computation.  

The top five curves in Fig.~1 show one half oscillation in the
acoustic wave model. It is notable that the shock-like features pass
through each other with little interference, leaving quite smooth
energy distributions behind them.

Fig.~2 shows the effect of lowering the initial amplitude $B$ in
Eq.~(\ref{eq:53}) by a factor of ten and increasing the
integration time by a factor of ten. As expected from
Eq.~(\ref{eq:wga_standing}), the time to appearance of
significant departures from a linear acoustic oscillation scales
inversely with the initial density contrast. It might be noted also
that decreasing $B$ increases the wavelengths of
the small-scale oscillations. 

Fig.~3 shows the energy distribution in a numerical solution to
the wave equation in two spatial dimensions, for the purpose of
checking whether the single spatial dimension in the first two
examples significantly limits the nature of the departure from
acoustic oscillations. The initial conditions are $\p y/\p t=0$
and 
\beq
y = 1 + B[\sin (k_xx) + \sin (k_yy)]/\sqrt{2}.
\label{eq:2d}
\eeq
The last factor makes the rms density contrast the
same as in Eq~(\ref{eq:53}). The wavenumber on the horizonal axis 
in Fig.~3 is the same as in Fig.~1, and the wavenumber on the
vertical axis is 50/77 times the horizontal wavenumber. The
energy contours are plotted at time $20T$, where $T$ is the
acoustic oscillation period for the mode running in the
horizontal direction. This is the same time as for the energy
distribution at the top curve in Fig.~1, and it
amounts to 12.99 times the acoustic period in the vertical
direction. The amplitude $B$ is the same as in Fig.~1, and the  
time step is the same, but there are only $50\times 77$ space
steps. The energy distribution from the WGM is
similar, though the density maximum at the lower left is narrower
in the horizontal direction and the minimum at the upper right is
broader. 

Better spatial resolution in this two-dimensional example would
give a clearer picture of the failure of the acoustic wave model
but would require a more ambitious computation. One does see the
departure from acoustic wave oscillation, and it appears at
about the same time as in the one-dimensional examples. There
is no indication in this example that the extra dimension
allows qualitatively different behavior.  

In an application to a model for structure formation in cosmology
\cite{HP}, small-scale shock-like features the in $y$-field
energy distribution could appear well before density fluctuations
of interest to astronomy reach the Hubble length. Fig.~4 shows a
check of the effect on the acoustic wave model for the large-scale
energy distribution. The initial conditions are $\p y/\p t=0$
and 
\beq
y = 1 + B[\sin (kx) - \sin (8kx)].
\label{eq:noise}
\eeq
The parameters in the numerical integration are the same as in
Fig.~1.

At time $t=2T$, at two acoustic oscillation periods of the long 
wavelength mode in Fig.~4, the short wavelength mode has completed 16
oscillations and the cascade to shorter wavelengths is
commencing. At $t=5T$ the
large-scale energy distribution is still close to the original
sine wave, consistent with the acoustic wave model. This
decoupling from the nonlinear small-scale 
behavior is familiar from other fluid models, and might 
be expected here on roughly similar grounds. In a region
where the smoothed energy density is larger than average the
$y$-field generally is oscillating more rapidly than average.
That produces a gradient in the phase of the field oscillation
that produces a flux of energy away from this region. Local 
nonlinear features produce local fluctuations in the phase
gradient but have little effect on its mean or on the mean energy
flux density.

Finally, a travelling wave solution offers a clearer illustration
of the formation of shock-like features. The initial
condition for the one-dimensional sinusoidal travelling wave in
Fig.~5 is 
\beqa
y(x,t) &=& (1+\dot\phi )\eta (t +\phi ),  \nonumber \\ 
\dot\phi &=& B\cos k(x-t/\sqrt{3}),
\label{eq:travelling_wave}
\eeqa
where $\eta (t)$ is a solution for the quartic oscillator with
unit amplitude in Eqs.~(\ref{eq:eta_eq}) and (\ref{eq:averages}). 
Eq.~(\ref{eq:travelling_wave}) is a solution to the wave
Eq~(\ref{eq:scaled_wave_eq}}) if terms of order $B^2$ and $kB$
may be ignored. I use a
numerical solution for $\eta (t)$, with $\eta =1$ and $\dot\eta =
0$ at $x=0$ in Eq.~(\ref{eq:travelling_wave}). This phase choice  
does not matter because $\eta$ oscillates much more rapidly
than the energy distribution. The amplitude $B$ and the other
parameters in the computation are the same as in the standing
wave example in Fig.~1. The growing departure from an acoustic
oscillation looks very much like the development of a shock wave,
as in Eq.~(\ref{eq:wga_travelling}). The time to
development of large departures from the acoustic model is about
half that of the standing wave, as might be expected from the
larger correction term in Eq.~(\ref{eq:wga_travelling}) and  
the coherence of phases of the leading and correction terms in
the travelling wave. The length of the train of oscillations
behind the shock-like feature doubles from time $10T$ to $20T$.
At $40T$ short wavelength oscillations in the energy density in
the numerical solution fill space, in a complicated pattern of  
amplitude and wavelength fluctuations. This part of the evolution
is not to be trusted, however, because the details are sensitive
to the size of the space step in the computation. 

\section{Concluding Remarks}

The discussion in Sec.~II shows that under the initial
conditions assumed here the initial departure from a homogeneous
energy distribution starts to
rearrange itself in the manner of an ideal relativistic fluid
with pressure $p_r=\rho _r/3$. Not so evident is that as the 
phase of the field oscillation as a function of position winds
and unwinds the energy distribution continues to behave like
a fluid. This behavior follows in the weak  
gradient model in Sec.~III (Eq.~\ref{eq:wga3}). The weak gradient
model in turn is equivalent to the equation of motion of a
relativistic fluid to second order in the 
perturbation from a homogeneous energy distribution, provided the
flow is irrotational in first order perturbation theory. 

In the numerical tests in Figs.~1 to~5 the weak gradient model
agrees well with the energy distribution obtained from numerical
solutions of the wave equation. That is, we have good evidence
for the picture of how the near acoustic wave behavior of the
energy distribution ends: features that resemble shock waves
appear after a time on the order of  
\beq
\tau _{\rm nl}\sim T/\delta\rho /\rho ,
\eeq
where $T$ is the oscillation time in the acoustic model 
and the rms density contrast is $\delta\rho /\rho$. 

The features are not true shock waves, of course. In particular,
they excite energy fluctuations in a 
reversible way: the field returns to a smooth energy distribution
after the feature has passed by. There is no artificial viscosity 
in the computation, and I have not been able to see how the effect
could be a numerical artifact; this seems to be real reversible
process. The near elastic interactions illustrated in the 
half cycle in Fig.~1 remind one of solitions, but this behavior
has a limited lifetime: the length of the train of oscillations
behind the leading edge increases with time. 

More ambitious analyses than have been attempted here would be
needed to explore the evolution of the shock-like features. If,
as seems likely, the 
fluctuations in the spatial distribution of the field energy
eventually cascade to noise extending to wavenumbers comparable
to the frequency $\omega$ of the field oscillation  
(Eq.~\ref{eq:omega}), the energy distribution may be expected to
end up behaving like a gas of interacting particles with
de~Broglie wavenumber on the order of $\omega$ \cite{Gruzinov}.
The near fluid behavior on scales much larger than the de~Broglie
wavelength might still obtain, however, because the gas has a
short mean free path. This aspect of the evolution remains to be
analyzed.   

\section{Acknowledgments}

This work has benefitted from discussions with Andrei Gruzinov,
Wayne Hu, and Alex Vilenkin, and was 
supported in part at the Institute for Advanced Study
by the Alfred P. Sloan Foundation and at Princeton University by
the National Science Foundation.

\begin{figure}
\centerline{\psfig{file=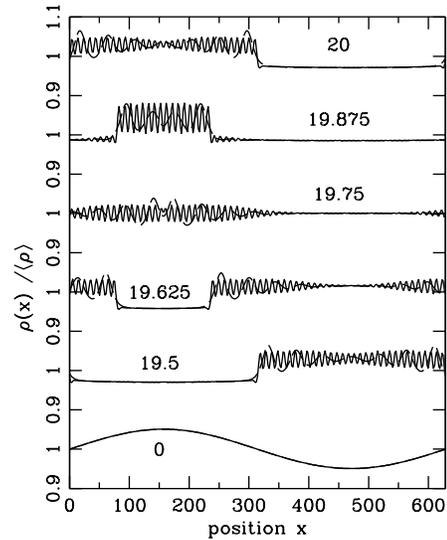,width=3truein,clip=}}
\caption{Evolution of a one-dimensional standing wave in the
energy distribution of a field with a quartic self-interaction
potential. The solid lines are from a numerical integration of
the wave equation (Eq.~\ref{eq:scaled_wave_eq}), and the dashed
lines are a numerical solution of the weak gradient model
(Eq.~\ref{eq:wga}). The energy distributions are plotted at the
indicated multiples of the period of oscillation of the initial
sine wave in the linear acoustic wave model.}  
\end{figure} 

\begin{figure}
\centerline{\psfig{file=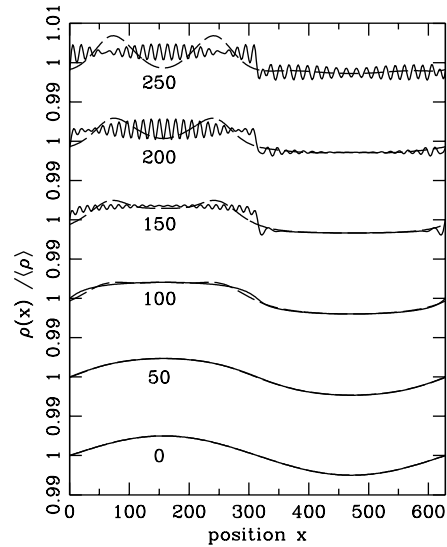,width=3truein,clip=}}
\caption{Illustration of the effect of decreasing the
wave amplitude in Fig.~1 by a factor of ten. The curves are
plotted at the indicated multiples of the acoustic oscillation
time. All other parameters are the same as in Fig.~1.}
\end{figure}

\begin{figure}
\centerline{\psfig{file=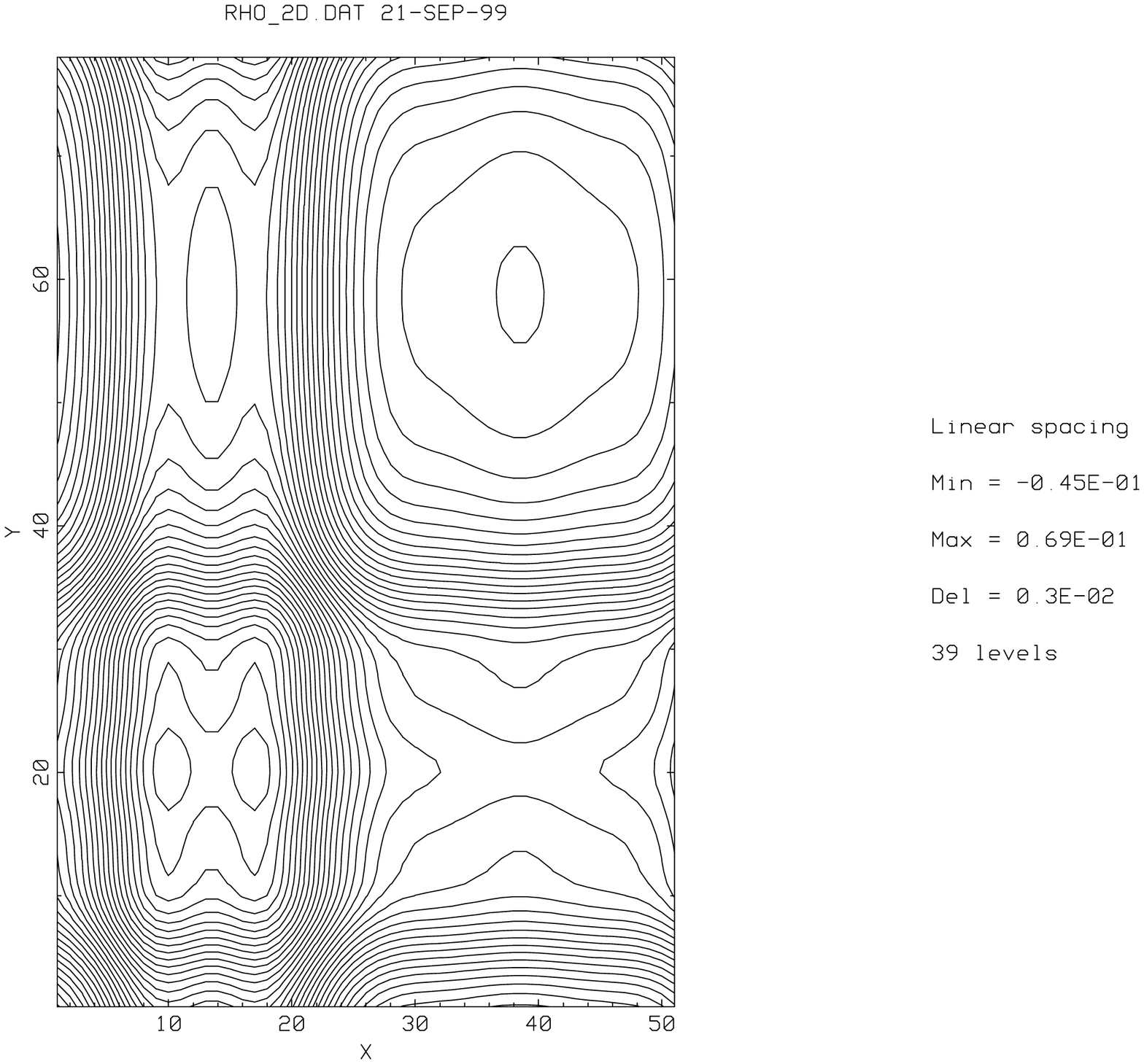,width=2truein,angle=270,clip=}}
\caption{The two-dimensional energy distribution after 20 times
the period for an acoustic oscillation in the horizontal 
direction. This is 13 times the acoustic oscillation
period in the vertical direction. The energy distribution is from
a numerical integration of the field equation with the initial
condition in Eq.~(\ref{eq:2d}). The time step, amplitude
parameter $B$, and wavenumber in the horizontal direction are
the same as in Fig.~1, but the spatial resolution is worse by a
factor of seven. The highest density contours, in the two 
peaks at the lower left, are at contrast 
$\delta\rho /\rho =0.069$; the lowest contour, at 
$\delta\rho /\rho =-0.045$, is at the upper right. The nearly
flat regions in the upper left and lower right are at close to
zero density contrast.}
\end{figure}

\begin{figure}
\centerline{\psfig{file=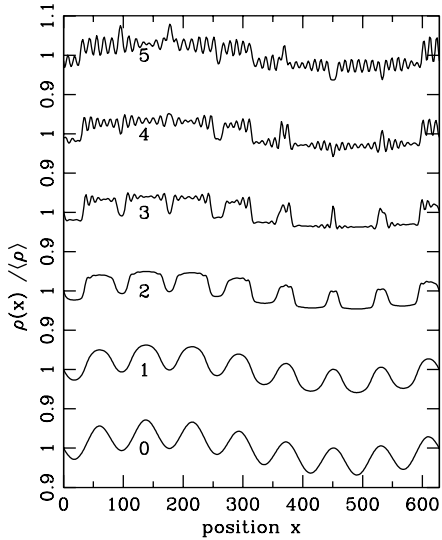,width=3truein,clip=}}
\caption{Illustration of the effect of small-scale nonlinear
fluctuations on evolution of the mass distribution on larger
scales. The initial condition is in Eq.~(\ref{eq:noise}). 
At the time of the top curve the long wavelength component has
oscillated five times and the short wavelength component 40 times
in the acoustic wave model. The other parameters are the same as
in Fig.~1.} 
\end{figure}

\begin{figure}
\centerline{\psfig{file=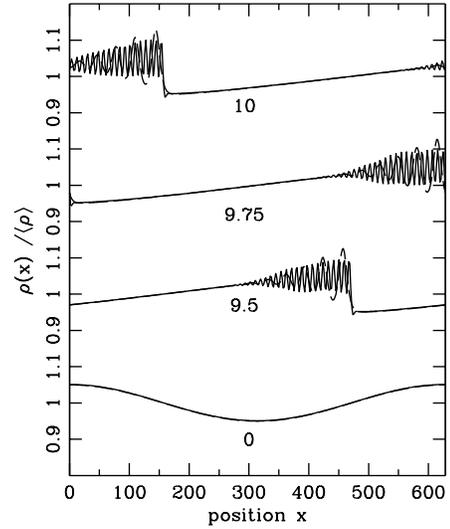,width=3truein,clip=}}
\caption{A travelling wave solution with the initial condition in 
Eq.~(\ref{eq:travelling_wave}). All other parameters are the same
as in Fig.~1.}
\end{figure}

\end{document}